\documentclass[aps,prx,twocolumn,floatfix,longbibliography,superscriptaddress]{revtex4-2}

\usepackage{amsmath}
\usepackage{amssymb}
\usepackage{times}
\usepackage{braket}
\usepackage[pdftex]{graphicx}
\usepackage{color}
\usepackage{blindtext}
\usepackage{nicefrac}
\usepackage[colorlinks=true, citecolor=blue, urlcolor=blue, linkcolor=blue]{hyperref}

\usepackage{todonotes}

\renewcommand{\vec}[1]{\boldsymbol{#1}}

\newcommand{\change}[1]{\textcolor{black}{#1}}

\renewcommand{\ket}[1]{\lvert#1\rangle} 
\newcommand{\braopket}[3]{\langle #1 | #2 | #3\rangle} 

\begin{document}

\title{Three-dimensional topological orbital Hall effect caused by magnetic hopfions}

\author{B{\"o}rge G{\"o}bel}
\email[Correspondence email address: ]{boerge.goebel@physik.uni-halle.de}
\affiliation{Institut f\"ur Physik, Martin-Luther-Universit\"at Halle-Wittenberg, D-06099 Halle (Saale), Germany}

\author{Samir Lounis}
\affiliation{Institut f\"ur Physik, Martin-Luther-Universit\"at Halle-Wittenberg, D-06099 Halle (Saale), Germany}

\date{\today}

\begin{abstract}
Magnetic hopfions are non-collinear spin textures that are characterized by an integer topological invariant, called Hopf index. The three-dimensional magnetic solitons can be thought of as a tube with a twisted magnetization that has been closed at both ends to form a torus. The tube consists of a magnetic whirl called in-plane skyrmion or bimeron. Although hopfions have been observed by microscopy techniques, their detection remains challenging as they lack an electronic hallmark so far. Here we predict a three-dimensional orbital Hall effect caused by hopfion textures: When an electric field is applied, the hopfion generates a transverse current of orbital angular momentum. The effect arises due to the local emergent field that gives rise to in-plane and out-of-plane orbital Hall conductivities. This orbital Hall response can be seen as a hallmark of hopfions and allows us to distinguish them from other textures, like skyrmioniums, that look similar in real-space microscopy experiments. While the two-dimensional topological invariant of a skyrmion determines its topological Hall transport, the unique three-dimensional topological orbital Hall effect can be identified with the three-dimensional topological invariant that is the Hopf index. Our results make hopfions attractive for spin-orbitronic applications because their orbital signatures allow for their detection in devices and give rise to large orbital torques.
\end{abstract}

\maketitle


\section{Introduction}

Non-collinear spin textures have emerged as a central topic in condensed matter physics due to their rich physics and potential applications in spintronic devices. Unlike collinear magnetic systems, where spins align uniformly in parallel or antiparallel configurations, non-collinear arrangements exhibit spatially varying spin orientations that can even give rise to topologically non-trivial spin textures, such as the family of two-dimensional magnetic whirls~\cite{gobel2021beyond} including skyrmions~\cite{bogdanov1989thermodynamically,muhlbauer2009skyrmion,yu2010real,nagaosa2010anomalous}, antiskyrmions~\cite{nayak2017magnetic,peng2020controlled,jena2020elliptical} and bimerons~\cite{kharkov2017bound,gobel2018magnetic,gao2019creation}. These textures can be considered as solitons that can be moved by current-induced spin torques and can be easily identified by real-space microscopy imaging techniques. Their topological nature gives rise to an effective magnetic field, called emergent field~\cite{nagaosa2013topological}, that causes a topological Hall effect which is a unique signature of skyrmions~\cite{bruno2004topological,neubauer2009topological,lee2009unusual,raju2021colossal} and many related textures~\cite{gobel2017afmskx,gobel2018family,gobel2018magnetic,goebel2019electrical,sivakumar2020topological}.

\begin{figure*}[t!]
    \centering
    \includegraphics[width=1\textwidth]{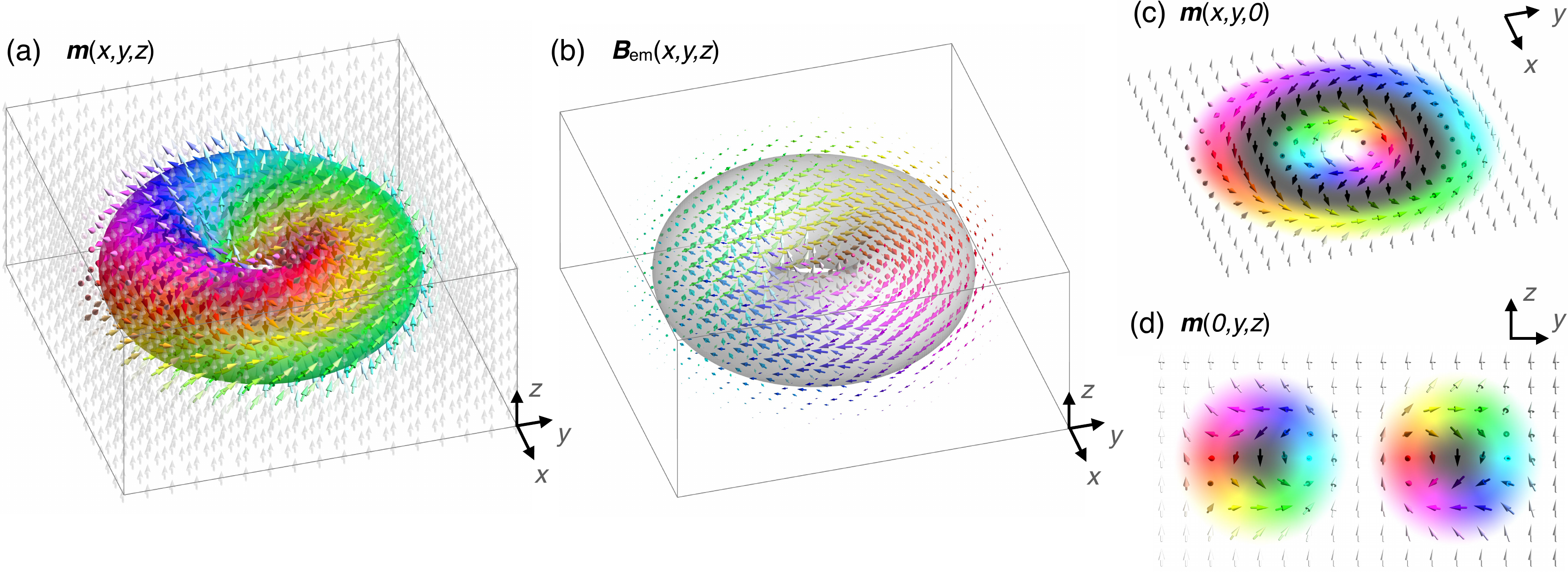}
    \caption{Magnetic hopfion. (a) Magnetic texture $\vec{m}(\vec{r})$ of a $\lambda=8a$ hopfion. (b) Corresponding emergent field $\vec{B}_\mathrm{em}(\vec{r})$. (c) A cut along the $xy$ plane reveals a magnetic skyrmionium. (d) A cut along the the $yz$ plane reveals two in-plane skyrmions also called bimerons. The color of the arrows in all panels encodes their orientation in the $xy$ plane. White and black arrows point along the positive and negative $z$ direction, respectively. \change{Note that these arrows that surround the hopfion have been plotted translucently to allow for a clearer view of the non-collinear part of the texture.}}
    \label{fig:overview}
\end{figure*}

Magnetic hopfions~\cite{hopf1964abbildungen,korepin1975quantization,sutcliffe2007vortex,sutcliffe2017skyrmion,sutcliffe2018hopfions,liu2018binding,tai2018static,wang2019current,gobel2020topological,liu2020three,raftrey2021field,kent2021creation,rybakov2022magnetic,popadiuk2023emergent,zheng2023hopfion} consist of a bimeron tube that has been twisted once and that has been merged at both ends to form a torus. These non-collinear magnetic solitons are innately three-dimensional [cf. Fig.~\ref{fig:overview}(a)], which makes their physics even richer as they possess more degrees of freedom. For example, not only can they be moved by currents~\cite{wang2019current,liu2020three} but also tilted~\cite{gobel2020topological,liu2020three}. Detecting these objects remains a difficult challenge, since hopfions do not exhibit a sizable topological Hall effect~\cite{gobel2020topological} and typical real-space microscopy techniques rely on two-dimensional projections of the magnetic textures, which is problematic for a three-dimensional soliton: The detectable signals caused by hopfions look similar to those of a skyrmionium~\cite{zhang2016control,zhang2018real,goebel2019electrical} -- a skyrmion that resides at the center of another skyrmion with reversed magnetization. Still, the experimental observation of hopfions has been achieved using a combination of X-ray photoemission electron microscopy (X-PEEM) and magnetic soft X-ray transmission microscopy (MTXM)~\cite{kent2021creation} or Lorentz transmission electron microscopy (LTEM)~\cite{zheng2023hopfion}. However, a more conclusive detection method would reveal the actual three-dimensional texture which is, in principle, possible by holography or tomography techniques using X-rays or electrons~\cite{wolf2019holographic,midgley2009electron,hierro2020revealing,wolf2022unveiling,seki2022direct,yasin2024bloch,winterott2025unlocking,gubbiotti20242025}. Unfortunately, these techniques are highly demanding and not compatible with potential spintronic applications of hopfions. A simple electric or magnetic signature of hopfions is desired to make these textures usable in devices.

Therefore, we consider the research field of orbitronics which has attracted great interest recently. It is concerned with the orbital angular momentum and the related orbital current. In equilibrium, the orbital angular momentum of a solid is often quenched because of the high symmetry, which is why the magnetization of most ferromagnets originates from the spin degree of freedom. However, when an electric field drives a system out of equilibrium, a sizable density of orbital magnetic moment or transverse orbital currents can be generated. These phenomena are called orbital Edelstein effect~\cite{levitov1985magnetoelectric,yoda2015current,yoda2018orbital, go2017toward,salemi2019orbitally,johansson2021spin,liu2021chirality,kim2023optoelectronic,el2023observation,hagiwara2024orbital,lee2024orbital,gobel2025chirality,gobel2025chirality2,busch2025non} and orbital Hall effect~\cite{zhang2005intrinsic, bernevig2005orbitronics, kontani2008giant, tanaka2008intrinsic, kontani2009giant,go2018intrinsic, pezo2022orbital,canonico2020orbital,cysne2022orbital,salemi2022theory,busch2023orbital,choi2023observation,lyalin2023magneto,busch2024ultrafast,gobel2024OHE,gobel2024topological}, respectively, and they are often larger than their spin counter-parts. The orbital effects often exist even without the relativistic spin-orbit coupling and only a small portion of these orbital signals gets converted to a spin Edelstein effect or a spin Hall effect due to this relativistic mechanism. Non-collinear spin textures with a non-trivial topology, such as the magnetic skyrmion, give rise to an orbital magnetization~\cite{dos2016chirality,lux2018engineering,gobel2018magnetoelectric} in equilibrium, and have been shown to exhibit a topological orbital Hall effect once an electric field is applied~\cite{gobel2024topological}. Even textures with a compensated emergent field, such as antiferromagnetic skyrmions~\cite{barker2016static,zhang2016magnetic,zhang2016antiferromagnetic,legrand2020room}, can exhibit such an effect~\cite{gobel2024topological} due to the local presence of an emergent field, even though it is compensated on average. 

In this paper, we use a tight-binding and Kubo approach and predict a three-dimensional topological orbital Hall effect caused by hopfion textures. This effect is generated by their local emergent field that gives rise to an orbital Berry curvature [cf. Fig.~\ref{fig:bandstructure}] and causes out-of-plane and in-plane orbital Hall conductivities [cf. Fig.~\ref{fig:hopfion7}]. The orbital Hall signatures of a hopfion are unique and allow us to distinguish it from two-dimensional topological textures, such as the skyrmion, bimeron or skyrmionium. While the two-dimensional topological invariant of a skyrmion determines its topological Hall transport, the unique three-dimensional orbital Hall transport is a manifestation of the three-dimensional topological invariant called Hopf index. Although all results presented in this paper rely on quantum mechanical calculations, they can be understood semiclassically on the basis of the emergent field of the textures and the carrier density of the conduction electrons.

This paper is structured as follows. We begin by detailing the tight-binding method used throughout this paper. In Sec.~\ref{sec:texture} we discuss the considered hopfion texture that enters the Hamiltonian and in Sec.~\ref{sec:conductivities} we present the Kubo approach that allows us to calculate the orbital Hall conductivity. The results of these calculations are presented in Sec.~\ref{sec:results_transport}. In Sec.~\ref{sec:results_comparison} we compare the results with the Hall transport caused by textures that are closely related to hopfions: 1. skyrmion tubes, 2. bimeron tubes and 3. skyrmionium tubes. The energy dependencies of the orbital Hall conductivities can be understood by analyzing the carrier density and the emergent magnetic field of these textures; cf. Sec.~\ref{sec:results_approximation}. Finally, we conclude in Sec.~\ref{sec:conclusion}.


\section{Model and Methods} \label{sec:model}

Throughout this paper, we use a tight-binding model to describe the conduction electrons and their spins' interaction with the hopfion texture (Sec.~\ref{sec:texture}). The orbital conductivity tensor elements are calculated based on a Kubo approach (Sec.~\ref{sec:conductivities}).
\subsection{Hopfion texture} \label{sec:texture}

First of all, we want to introduce the hopfion texture $\vec{m}(\vec{r})$ and discuss its topological properties. $\vec{m}$ is the normalized magnetic moment and $\vec{r}$ is the position vector. We use the top-down view coordinates $x,\,y$ with $r=\sqrt{x^2+y^2}$ the distance and $\phi=\arctan(y/x)$ the polar angle, as well as the transformed coordinates $\rho_x=r-\lambda/2,\,\rho_y=z$ 
with $\rho=\sqrt{\rho_x^2+\rho_y^2}$. The hopfion is modeled as
\begin{align}
    \vec{m}(\vec{r})=\begin{pmatrix}
        \sin(2\pi\rho/\lambda)\frac{x\rho_y-y\rho_x}{r\rho}\\
        \sin(2\pi\rho/\lambda)\frac{x\rho_x+y\rho_y}{r\rho}\\
        -\cos(2\pi\rho/\lambda)
    \end{pmatrix}\label{eq:hopfion}
\end{align}
in the volume of the torus $\sqrt{(x-\lambda/2\cdot\cos\phi)^2+(y-\lambda/2\cdot\sin\phi)^2+z^2}<\lambda/2$ and $\vec{m}(\vec{r})=\vec{e}_z$ elsewhere. This texture is topologically equivalent to the ones used in the literature, e.\,g. Ref.~\cite{sutcliffe2018hopfions,gobel2020topological}, but has a homogeneous background. This is important because the orbital Hall transport will be calculated based on a reciprocal-space approach, so we need periodicity in real space. The texture above describes the texture in a unit cell of size $2\lambda\times2\lambda\times\lambda$. 

The hopfion texture is shown in Fig.~\ref{fig:overview}(a) and can be thought of as a skyrmion tube that is magnetized in the plane of the tube, has been twisted once, and whose ends have been merged to form a torus. In our calculations, the torus is located in the $xy$ plane. 
If the three-dimensional magnetic texture is cut along the $xy$ plane, the result is a skyrmionium -- a skyrmion that is positioned at the center of another skyrmion with opposite magnetization. This cut containing the skyrmionium is shown in Fig.~\ref{fig:overview}(c). If one cuts the hopfion along a plane that contains the $z$ axis, e.\,g. the $yz$ plane, one can see the texture of the skyrmion tube that forms the hopfion. However, as explained before, it is actually an in-plane magnetized skyrmion, which is called bimeron, and since the torus is cut twice by the plane, we see two bimerons in Fig.~\ref{fig:overview}(d). In Sec.~\ref{sec:results_comparison} we will show that the orbital Hall response of a hopfion is related to the orbital Hall responses of these two-dimensional textures.

An explanation will be provided on the basis of the emergent field of a magnetic texture that is a quantity that determines its topological Hall transport and orbital Hall transport. Its $\alpha=x,y,z$ component is defined as~\cite{nagaosa2013topological}
\begin{align}
B_{\mathrm{em},\alpha}(\vec{r})=\frac{1}{2}\sum_{\beta,\gamma}\epsilon_{\alpha\beta\gamma}\,\vec{m}(\vec{r})\cdot\left(\frac{\partial\vec{m}(\vec{r})}{\partial \beta}\times \frac{\partial\vec{m}(\vec{r})}{\partial \gamma}\right).
\end{align}
The emergent field of a hopfion is shown in Fig.~\ref{fig:overview}(b). Unlike for the two-dimensional textures discussed before, it is non-collinear and has a toroidal configuration in the hopfion torus. In the center of the hopfion it points out of the torus plane. On average, the emergent field of the hopfion is compensated.

Two-dimensional topological textures, like magnetic skyrmions, bimerons and skyrmioniums that constitute cuts of the hopfion are characterized by the topological charge
\begin{align}
N_\mathrm{Sk}=\frac{1}{4\pi}\int\vec{m}(\vec{r})\cdot\left(\frac{\partial\vec{m}(\vec{r})}{\partial \beta}\times \frac{\partial\vec{m}(\vec{r})}{\partial \gamma}\right)\,\mathrm{d}^2r.
\end{align}
Here we have assumed that the texture changes in the $\beta\gamma$ plane and is constant along $\alpha$. In this case, the topological charge $N_\mathrm{Sk}$ is the integral of the perpendicular emergent field $B_\mathrm{em,\alpha}(\vec{r})$ over the whole texture. While a skyrmion and a bimeron tube are characterized by a non-trivial topological invariant $N_\mathrm{\mathrm{Sk}}=\pm1$, the two-dimensional cuts through the hopfion always contain two skyrmions or two bimerons with opposite topological charges. Therefore, the skyrmionium in Fig.~\ref{fig:overview}(c) and the pair of bimerons in Fig.~\ref{fig:overview}(d) are topologically trivial, $N_\mathrm{Sk}=0$.

A hopfion is a three-dimensional texture. Therefore, it is not reasonable to characterize it by the topological charge $N_\mathrm{Sk}$ as defined above. Still, the emergent field is important to characterize its topology. A hopfion is characterized by the Hopf index~\cite{whitehead1947expression,wilczek1983linking,knapman2025numerical}
\begin{align}
N_H=-\frac{1}{(4\pi)^2}\int\vec{B}_\mathrm{em}(\vec{r})\cdot\vec{A}(\vec{r})\,\mathrm{d}^3r.
\end{align}
It is $N_H=\pm 1$ depending on the type of hopfion and is calculated from the emergent field $\vec{B}_\mathrm{em}(\vec{r})$ and the corresponding vector potential $\vec{A}(\vec{r})$. Like every vector potential it is not gauge invariant and fulfills $\nabla\times\vec{A}=\vec{B}_\mathrm{em}$. The emergent field and the Hopf index are gauge-invariant. So far, the role of the Hopf index has only been discussed in terms of topological stability of the hopfion but not in terms of transport properties. This is the purpose of this work as we identify the unique orbital Hall response caused by hopfions.

Although the textures as defined throughout this section are continuous functions of the position vector $\vec{r}$, for the tight-binding calculations we consider a discrete lattice. Therefore, the texture $\vec{m}(\vec{r})$ is rather a set of magnetic moments $\{\vec{m}_i\}_{i=1,...,N}$ that are defined at the sites $\{\vec{r}_i\}_{i=1,...,N}$. Since the calculation of a three-dimensional magnetic texture is computationally demanding, we are restricted when choosing appropriate hopfion sizes for the calculations. Here we use a unit cell consisting of $N=256$ magnetic moments which is enough to capture the topological properties of the hopfion as well as of the skyrmionium and bimerons accurately.


\subsection{Calculation of Hall conductivities} \label{sec:conductivities}

The electronic Hamiltonian is based on a tight-binding approach and is written here in second quantization~\cite{ohgushi2000spin,hamamoto2015quantized,gobel2017THEskyrmion,yin2015topological,gobel2024topological}
\begin{align} 
  H &  =  -t \sum_{\braket{ij}} \,c_{i}^\dagger c_{j} + m \sum_{i} \vec{m}_{i} \cdot (c_{i}^\dagger \boldsymbol{\sigma}c_{i}).
  \label{eq:ham_the} 
\end{align} 
$c_{i}^\dagger$ and $c_{i}$ are the creation and annihilation operators of an $s$ electron at site $i$. $\vec{\sigma}$ is the vector of Pauli matrices. \change{The Hamiltonian has the dimension $2N\times2N$.}

\begin{figure}[t!]
    \centering
    \includegraphics[width=\columnwidth]{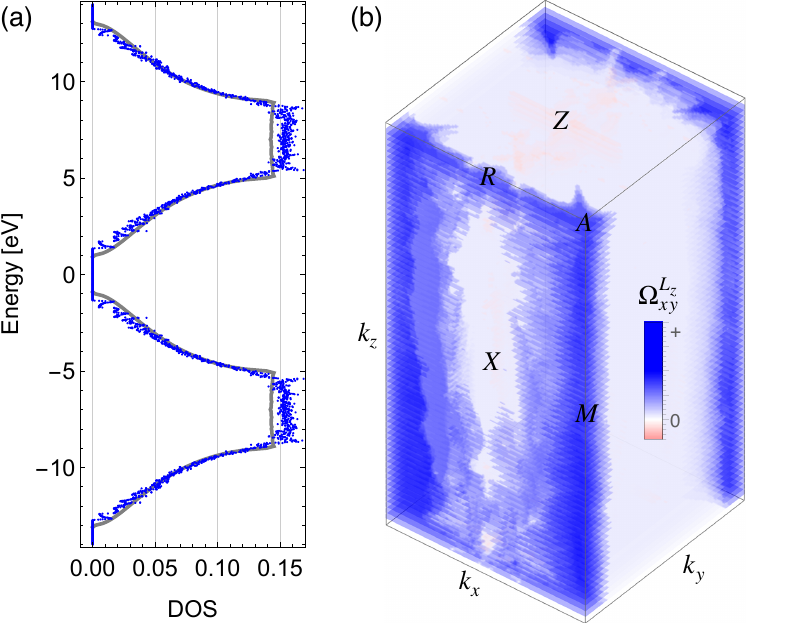}
    \caption{Electronic properties caused by a hopfion with $m=7t$. (a) Density of states \change{[Eq.~\eqref{eq:dos}]} (blue) and comparison to the density of states without the hopfion (gray). (b) Slices of the orbital Berry curvature $\Omega_{\nu,xy}^{L_z}(\vec{k})$ of the lowest band $\nu=1$.}
    \label{fig:bandstructure}
\end{figure}

The first term describes the kinetic energy of the conduction electrons via hopping terms \change{of the form $-t\exp[i\vec{k}\cdot(\vec{r}_i-\vec{r}_j)]$. Here, $\vec{k}$ is the wave vector and $\vec{r}_i-\vec{r}_j$ the hopping path.} For the hopping amplitude we choose $t=1\,\mathrm{eV}$ and for the lattice constant $a=2.76\,\mathrm{\AA}$. We consider a three-dimensional cubic lattice. The second term describes the coupling of the conduction electrons with the magnetic texture $\{\vec{m}_i\}$ that is formed by energetically lower states; e.\,g. by $d$ electrons. 
Throughout this paper, we consider two cases: (i) the adiabatic limit where the conduction electrons' spins align almost perfectly with the texture (here $m=7t$), and the weak-coupling case, where spin parallel and antiparallel states are hybridized (here $m=2t$).

\change{Eq.~\eqref{eq:ham_the} is the so-called $sd$-Hamiltonian in which only $s$ electrons are considered explicitly and $d$ electrons enter only via their magnetic texture $\{\vec{m}_i\}$. This Hamiltonian is suitable for analyzing topological transport phenomena~\cite{ohgushi2000spin,hamamoto2015quantized,gobel2017THEskyrmion,yin2015topological,gobel2024topological} and has the advantage that spin-orbit coupling is zero, so that the pure topological influence of the magnetic texture on the Hall conductivities can be calculated. Note that spin-orbit coupling is not required for the emergence of a topological Hall effect, topological orbital Hall effect or topological spin Hall effect~\cite{gobel2024topological}.}

We diagonalize the $sd$-Hamiltonian to \change{numerically} determine its eigensystem. The eigenvalues $E_{\nu\vec{k}}$ constitute the band structure with band index \change{$\nu=1,...,2N$}. From the eigenvectors $\ket{\nu\vec{k}}$ we calculate the Hall conductivities. 
The orbital Hall transport is quantified by the orbital Hall conductivity~\cite{pezo2022orbital}
\begin{align}
    \sigma^{L_z}_{xy}(E_\text{F})= \frac{e}{\hbar}\sum_\nu \frac{1}{(2\pi)^3}\int_{E_{\nu \vec{k}}\leq E_\text{F}}\Omega_{\nu,xy}^{L_z}(\vec{k}) \,\mathrm{d}^3k.\label{EQ:sigma_Lz_xy_Kubo}
\end{align}
This equation assumes temperature $T=0$ and is a function of the Fermi energy $E_\mathrm{F}$. We integrate over the orbital Berry curvature 
\begin{align}
    \Omega_{\nu,xy}^{L_z}(\vec{k})&= -2 \hbar^2\ \text{Im}\ \sum_{\mu\neq \nu} \frac{\braopket{\nu \vec{k}}{j_x^{L_z}}{\mu \vec{k}} \braopket{\mu \vec{k}}{v_y}{\nu \vec{k}}}{(E_{\nu \vec{k}} - E_{\mu \vec{k}})^2}.
\end{align}
\change{Although the electronic transport is determined by all occupied states, the influence of the $d$ electrons is negligible because they are fully occupied and far away from the $s$ states near the Fermi level.} For calculating the orbital Hall conductivity at finite temperatures $T>0$, the orbital Berry curvature is weighted by the Fermi-Dirac distribution function in the energy integral. 

We use the velocity operator $v_l=\frac{1}{\hbar}\frac{\partial H}{\partial k_l}$ and the orbital current operator $\braopket{\nu \vec{k}}{j_x^{L_z}}{\mu \vec{k}} = \frac{1}{2}\sum_{\alpha} [ \braopket{\nu \vec{k}}{v_x}{\alpha \vec{k}} \braopket{\alpha \vec{k}}{L_z}{\mu \vec{k}} + \braopket{\nu \vec{k}}{L_z}{\alpha \vec{k}} \braopket{\alpha \vec{k}}{v_x}{\mu \vec{k}} ]$. The orbital angular momentum
\begin{align}
      &\braopket{\nu \vec{k}}{L_z}{\alpha \vec{k}} = i  \frac{e\hbar^2}{4g_L\mu_\mathrm{B}}  \sum_{\beta \neq \nu, \alpha} \left( \frac{1}{E_{\beta \vec{k}} - E_{\nu \vec{k}}} + \frac{1}{E_{\beta \vec{k}} - E_{\alpha \vec{k}}} \right)\notag\\
      &\times\left(\braopket{\nu \vec{k}}{v_x}{\beta \vec{k}} \braopket{\beta \vec{k}}{v_y}{\alpha \vec{k}} - \braopket{\nu \vec{k}}{v_y}{\beta \vec{k}} \braopket{\beta \vec{k}}{v_x}{\alpha \vec{k}}\right) \label{eq:oam}
\end{align}
has been calculated based on the modern formulation~\cite{chang1996berry,oppeneer1998magneto,xiao2005berry,thonhauser2005orbital,ceresoli2006orbital,raoux2015orbital} and takes into account inter-site contributions~\cite{pezo2022orbital} that are not considered in the often used atomic-center approximation. Using this more accurate description has been shown to be of crucial importance for topologically non-trivial spin textures such as skyrmions, as the orbital angular momentum and the corresponding orbital Hall conductivity are caused by large cyclotron orbits generated by the texture's emergent magnetic field~\cite{gobel2024topological}.

Although the focus of this paper is on orbital transport, we also briefly mention the topological Hall conductivity and spin Hall conductivity, as defined in Ref.~\cite{gobel2024topological}.


\section{Results and Discussion} \label{sec:results}

Next, we present the results of our calculations and begin with the orbital Hall signatures of hopfions in Sec.~\ref{sec:results_transport}. Afterwards, in Sec.~\ref{sec:results_comparison}, we compare them with the typical Hall signatures of related non-collinear textures: 1. skyrmion tubes, 2. bimeron tubes and 3. skyrmionium tubes. In Sec.~\ref{sec:results_approximation}, we present how the energy dependence of the orbital Hall conductivities can be understood by analyzing the carrier density and the emergent field of these textures.


\subsection{Orbital Hall transport of hopfions} \label{sec:results_transport}

Diagonalizing the Hamiltonian [Eq.~\eqref{eq:ham_the}] and using the hopfion texture [Eq.~\eqref{eq:hopfion}] results in the band structure describing electrons interacting with a periodic array of hopfions. In the strong-coupling case, where $m=7t$, the band structure exhibits 2 blocks consisting of 256 bands each. The conduction electron spins align locally parallel and anti-parallel with the hopfion texture in the two blocks, respectively. The blocks are centered around $\pm m$ and have a band width of almost $12t$. Due to the large number of bands, it is not helpful to look at the band structure directly, as the bands are very dense. 

\begin{figure}[t!]
    \centering
    \includegraphics[width=\columnwidth]{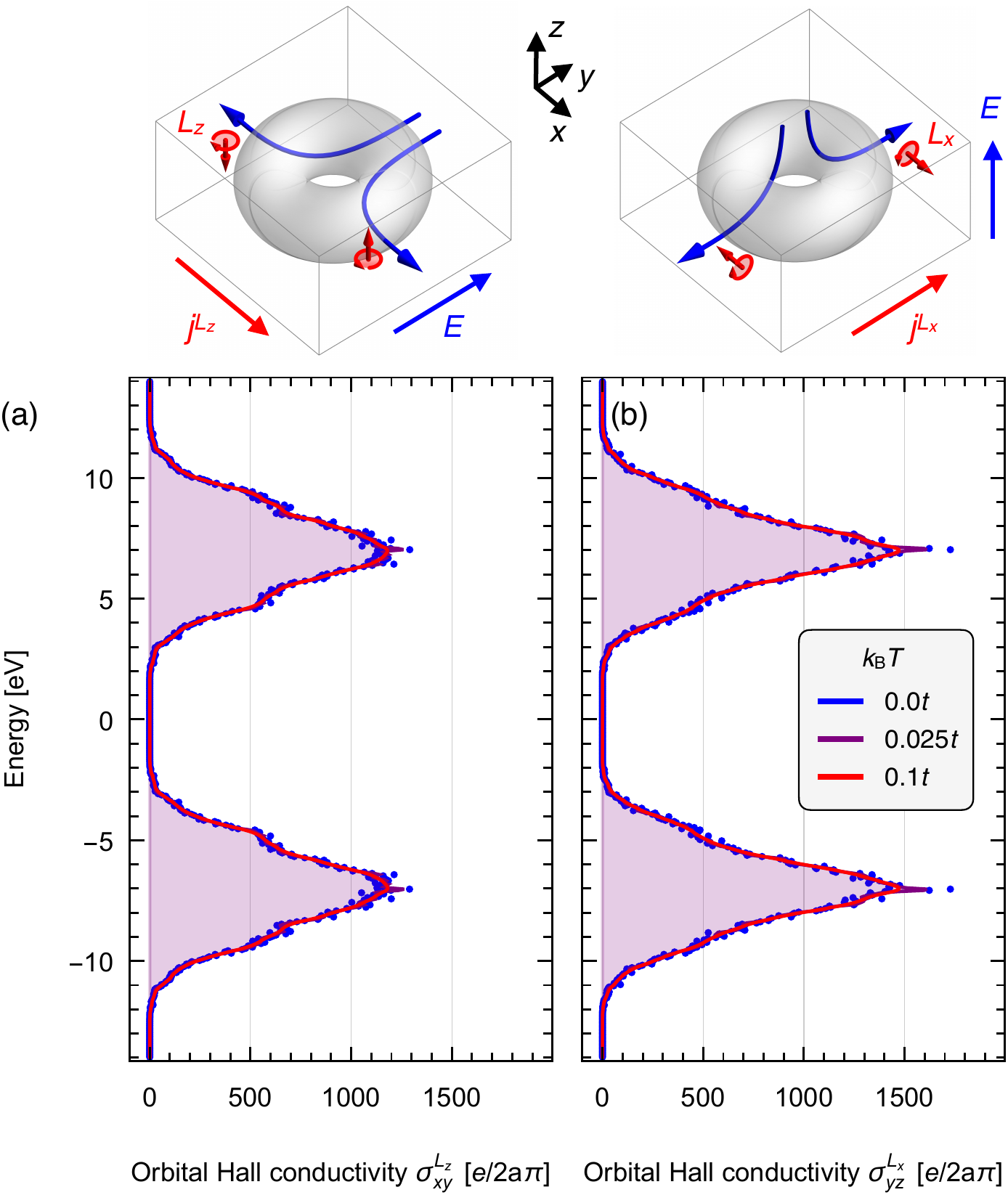}
    \caption{Three-dimensional orbital Hall effect caused by a hopfion in the strong-coupling limit. (a) The orbital conductivity tensor element $\sigma_{xy}^{L_z}$ as a function of energy. This orbital Hall conductivity characterizes a transport of orbital angular momentum $L_z$ in the Hopfion plane $xy$. Application of an electric field $\vec{E}$ along $y$ gives rise to an orbital current $\vec{j}^{L_z}$ along $x$; see schematic image above. The orbital angular momentum is polarized out of plane. (b) The equivalent results for the $\sigma_{yz}^{L_x}=\sigma_{zx}^{L_y}$ tensor element. The Hund's coupling is characterized by $m=7t$ so that the conduction electrons' spins almost completely align with the hopfion texture. The color of the curve indicates the temperature; see legend.}
    \label{fig:hopfion7}
\end{figure}

\subsubsection{Density of states}

Instead, the density of states, shown in Fig.~\ref{fig:bandstructure}(a) for the strong-coupling case $m=7t$, reveals better how the states are distributed in energy. \change{This function quantifies the states in an energy interval around $E$
\begin{align}
    \mathrm{DOS}(E)=\frac{1}{(2\pi)^3}\sum_\nu\int \delta(E-E_{\nu\vec{k}})\,\mathrm{d}^3k. \label{eq:dos}
\end{align}}

Near the band edges, we see an increase and a rather constant plateau near the centers of the blocks around $\pm m$. The density of states (blue) agrees rather well with the density of states of the band structure that characterizes the underlying square lattice (gray). It consists of 2 bands that we have shifted by $\pm m$
\begin{align}
    E_{3\mathrm{D}}(\vec{k})=-2t[\cos(k_xa)+\cos(k_ya)+\cos(k_za)]\pm m.
\end{align}
Due to the topological properties of the hopfion texture, an emergent field arises that acts just like an actual magnetic field that couples to the charge of the electrons. For a two-dimensional system, like a thin sample hosting skyrmions, it leads to the emergence of Landau levels whose energy is determined by the density of states~\cite{gobel2017QHE} according to Onsager's quantization scheme~\cite{onsager1952interpretation}. Even though here the discussion is more difficult, due to the three-dimensional nature of the system and the globally compensated emergent field, we still find that the density of states of the magnetic system (hosting a hopfion) and the non-magnetic system (characterized by $E_{3\mathrm{D}}(\vec{k})$) are relatable.

In summary, the band structure exhibits a dense collection of bands whose density of states agrees well with the density of states of the non-magnetic system. The two cosine-shaped bands of the cubic lattice condense to many rather flat bands because of the emergent field of the hopfion. Next, we analyze the orbital Hall conductivity that is determined by the eigenvectors that are affected by the emergent field as well.

\subsubsection{Orbital Hall transport}

The bands exhibit non-vanishing orbital angular momenta that have been calculated based on the modern formulation of orbital magnetization [Eq.~\eqref{eq:oam}]. Also, the hopfion gives rise to a net orbital Berry curvature [shown in Fig.~\ref{fig:bandstructure}(b) for the strong-coupling case $m=7t$] that causes an orbital Hall conductivity; cf. Fig.~\ref{fig:hopfion7}. Interestingly, since the emergent field [Fig.~\ref{fig:overview}(b)] has components in all directions in space, not only the orbital Hall tensor elements \change{$\sigma_{xy}^{L_z}=-\sigma_{yx}^{L_z}$} [presented in Fig.~\ref{fig:hopfion7}(a)] are non-zero but also the elements $\sigma_{yz}^{L_x}=-\sigma_{zy}^{L_x}=\sigma_{zx}^{L_y}=-\sigma_{xz}^{L_y}$ [presented in Fig.~\ref{fig:hopfion7}(b)]. This is unlike the orbital Hall effect of any two-dimensional texture in the $xy$ plane, which can only exhibit the orbital Hall conductivity \change{$\sigma_{xy}^{L_z}=-\sigma_{yx}^{L_z}$}, as discussed in more detail in Sec.~\ref{sec:results_comparison}. Therefore, we call this unique signature of the hopfion the three-dimensional orbital Hall effect.

The presence of an orbital Hall effect is not related to the finite size of the hopfion but can be expected even for a perfectly smooth texture because the orbital Hall conductivity is known to be insensitive to the sign of the emergent field~\cite{gobel2024topological}. For this reason, any texture with an emergent field can give rise to an orbital Hall effect even if it is compensated on average. We have recently demonstrated this for an antiferromagnetic skyrmion which consists of 2 sublattices with opposite emergent fields. These opposite fields cause the same orbital Hall conductivity because they give rise to opposite orbital angular momenta and opposite charge Hall transport~\cite{gobel2024topological}. 

The energy-dependent curves in Fig.~\ref{fig:hopfion7}(a,b) look rather similar for both tensor components except for a difference in magnitude. Furthermore, they are almost identical for the upper and lower block in energy and the signal is symmetric within each block, starting at zero near the band edges and exhibiting a maximum near the center of each block.
A detailed analysis of the energy dependence of these orbital conductivities and their relation to the emergent field follows later in Sec.~\ref{sec:results_approximation}. The magnitude of the orbital Hall conductivity scales with the component of $1/\braket{B_{\mathrm{em},\alpha}^2}$ that is perpendicular to the plane of transport. Here, the bracket indicates an average over the whole hopfion. The hopfion has a larger out-of-plane component $\braket{B_{\mathrm{em},z}^2}$ than the in-plane components $\braket{B_{\mathrm{em},x}^2}=\braket{B_{\mathrm{em},y}^2}$, which is why $\sigma_{yz}^{L_x}$ [presented in Fig.~\ref{fig:hopfion7}(b)] is larger than $\sigma_{xy}^{L_z}$ [presented in Fig.~\ref{fig:hopfion7}(a)] with maximum value of $\sim 1500\,\frac{e}{2a\pi}$ and $\sim 1200\,\frac{e}{2a\pi}$, respectively. 

\begin{figure}[t!]
    \centering
    \includegraphics[width=\columnwidth]{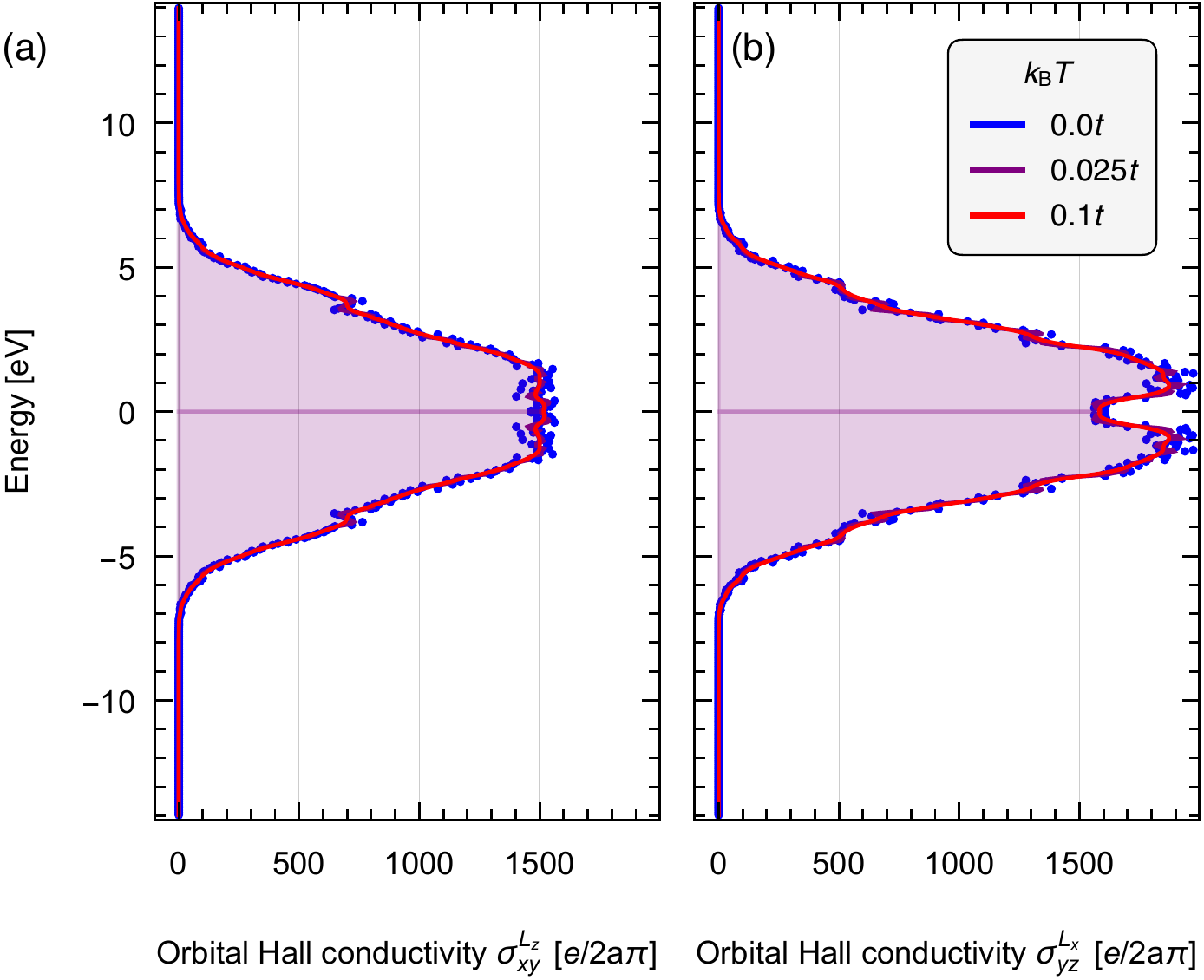}
    \caption{Three-dimensional orbital Hall effect caused by a hopfion in the weak-coupling limit. The figure is analogue to Fig.~\ref{fig:hopfion7} but with a weaker Hund's coupling of $m=2t$.  (a) The orbital conductivity tensor element $\sigma_{xy}^{L_z}$ and (b) the tensor element $\sigma_{yz}^{L_x}=\sigma_{zx}^{L_y}$ are shown as a function of energy for various temperatures.}
    \label{fig:hopfion2}
\end{figure}

The orbital Hall conductivity for the weak-coupling case, $m=2t$, is shown in Fig.~\ref{fig:hopfion2}. We see the same tensor components as before but there is no separation into two blocks anymore. Instead, spin parallel and antiparallel states that feel opposite emergent fields are hybridized. However, since opposite emergent fields result in the same orbital Hall response, this still leads to a considerable orbital Hall effect, even in the weak-coupling limit. In fact, for some energies, the orbital Hall conductivity is even larger for $m=2t$ than in the strong coupling case $m=7t$. This is because the density of states is larger when there is no formation of two separate spin-polarized blocks in the band structure.

\change{When comparing the in-plane orbital Hall conductivity $\sigma_{xy}^{L_z}$ in Fig.~\ref{fig:hopfion2}(a) with the out-of-plane orbital Hall conductivity $\sigma_{yz}^{L_x}$ in Fig.~\ref{fig:hopfion2}(b), we see similar energy dependencies but different peak structures near $E=0$. For a smooth and infinitely large texture with a homogeneous emergent field, we would expect qualitatively equal curves. However, because of the finite size of the hopfion, differences occur near the center of the energy range. Here electron-like spin parallel states hybridize with hole-like spin antiparallel states which is why small changes or inhomogeneities in the emergent field can lead to quite strong changes in the orbital Hall conductivities.}

\begin{figure*}[t!]
    \centering
    \includegraphics[width=\textwidth]{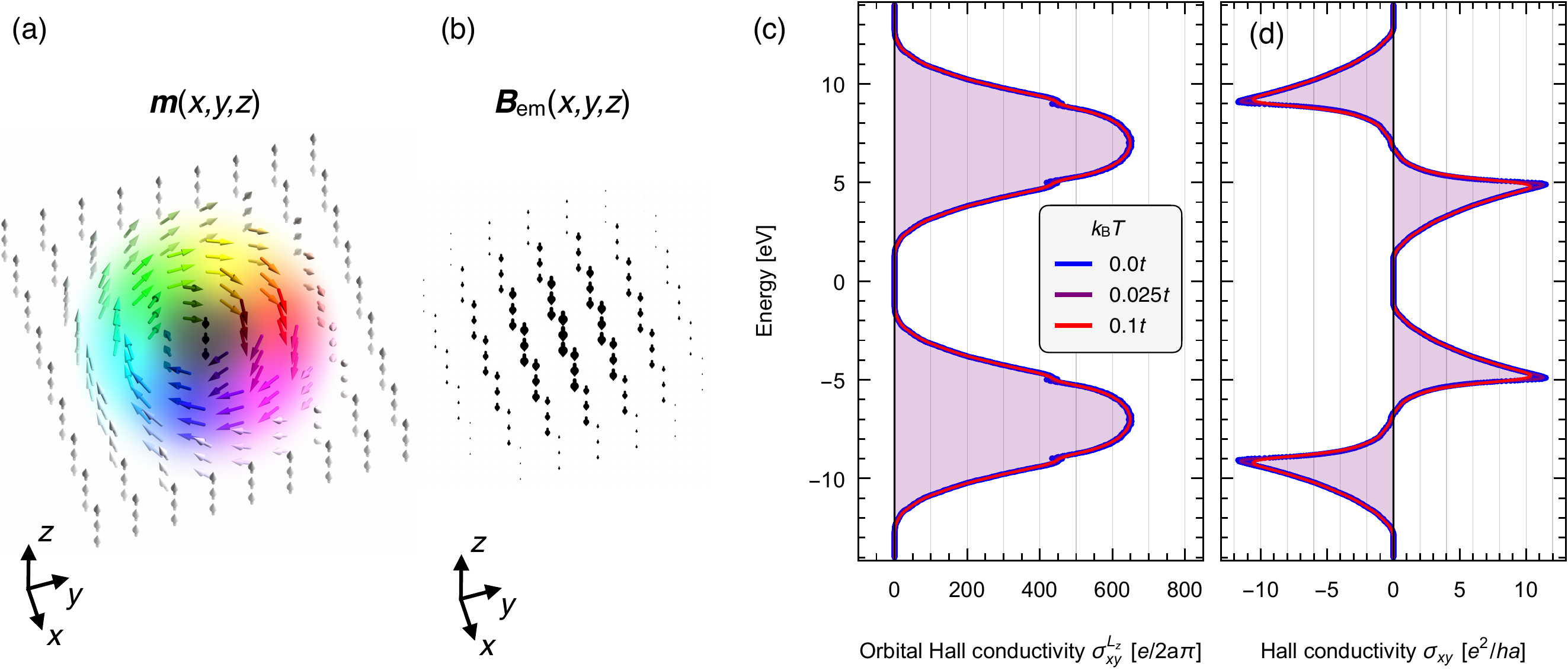}
    \caption{Orbital Hall effect and topological Hall effect caused by a skyrmion tube. (a) Magnetic texture of the skyrmion tube. The color indicates the orientation of the magnetic moments as in Fig.~\ref{fig:overview}. (b) Corresponding emergent field that is collinear and points along the tube direction. (c) Orbital Hall conductivity $\sigma_{xy}^{L_z}$ as a function of energy. (d) Topological Hall conductivity $\sigma_{xy}$ as a function of energy. The color of the curve indicates the temperature; see legend. The Hund's coupling is characterized by $m=7t$.}
    \label{fig:skyrmion}
\end{figure*}

So far, we have discussed only one type of hopfion. We have repeated our calculations from the strong-coupling limit, $m=7t$ presented in Fig.~\ref{fig:hopfion7}, for different types of hopfions. By reversing $m_z$, the Hopf index and the emergent field change sign. 
As a consequence, the band structure changes $E_\nu(k_x,k_y,k_z)\rightarrow E_\nu(k_x,k_y,-k_z)$.
However, all orbital Hall conductivity tensor elements remain unaffected by the change of the Hopf index.
The same happens when we reverse an in-plane component, say $m_x$, instead. Due to the opposite winding, the Hopf index and the emergent field change sign but the orbital Hall conductivity is unaffected. 
This fits well the explanation that the orbital Hall conductivity is caused by the square of the emergent field, which is not sign-sensitive. We will elaborate on this explanation in more detail in Sec.~\ref{sec:results_approximation}. 

Before we continue, we would like to note that the bands also exhibit a finite Berry curvature and spin Berry curvature that give rise to a small topological Hall effect and spin Hall effect, respectively. A smooth hopfion texture has a globally compensated emergent field and so these effects are not expected. In our calculations, they occur because of the finite size of the hopfion that is limited by the computation time. The hopfion texture is not exactly smooth, which is why we find a finite topological Hall and spin Hall conductivity near the band edges. However, these signals are weak when we compare them to other textures such as skyrmion tubes; cf. Sec.~\ref{sec:results_comparison}. The effects would disappear for larger hopfion sizes $\lambda/a\rightarrow\infty$.


\subsection{Comparison with related non-collinear textures} \label{sec:results_comparison}

Before we dissect the origin of the orbital Hall effect in more detail, first we want to compare our calculations with those for related non-collinear spin textures: 1. skyrmion tubes, 2. bimeron tubes and 3. skyrmionium tubes. We analyze their orbital Hall conductivities because a hopfion can be understood as a bimeron tube that has been twisted and closed to form a torus. This bimeron tube is topologically equivalent to a skyrmion tube, because a bimeron is essentially a skyrmion in an in-plane magnet. For this reason, cutting a hopfion along a plane containing the $z$ axis results in two bimerons; cf. Fig.~\ref{fig:overview}(d). When a hopfion is cut along the $xy$ plane instead, the resulting texture is a skyrmionium; cf. Fig.~\ref{fig:overview}(c). Therefore, the results for the bimeron and for the skyrmionium tube will help us to better understand the orbital Hall signatures of the hopfion.

\subsubsection{Skyrmion tube}

The first comparison with a two-dimensional texture that comes to mind is to relate our findings to the orbital Hall response of a skyrmion tube. This texture [Fig.~\ref{fig:skyrmion}(a)] exhibits a topologically non-trivial configuration in the two-dimensional $xy$ plane that is continued along the $z$ direction. The corresponding emergent field [Fig.~\ref{fig:skyrmion}(b)] is collinear and points along $z$ because the texture only changes along the $xy$ plane. 

We find a sizable \change{$\sigma_{xy}^{L_z}=-\sigma_{yx}^{L_z}$} orbital conductivity tensor element [Fig.~\ref{fig:skyrmion}(c) for the strong coupling case $m=7t$] that exhibits a similar energy dependence as for the hopfion; cf. with Fig.~\ref{fig:hopfion7}(a). The maximum is smaller with $\sim 650\,\frac{e}{2a\pi}$ which is due to the larger emergent field component $\braket{B_{\mathrm{em},z}^2}$ compared to the hopfion. The tensor elements $\sigma_{yz}^{L_x}$, $\sigma_{zy}^{L_x}$, $\sigma_{zx}^{L_y}$ and $\sigma_{xz}^{L_y}$ that were all finite for the hopfion are now zero due to $\braket{B_{\mathrm{em},x}^2}=\braket{B_{\mathrm{em},y}^2}=0$. Instead, we find a non-zero topological Hall conductivity $\sigma_{xy}=-\sigma_{yx}$, presented in Fig.~\ref{fig:skyrmion}(d), that was negligible for the hopfion. This is because the topolgical Hall effect scales with the average emergent field and $\braket{B_{\mathrm{em},z}}$ is finite for the skyrmion while it is compensated for the hopfion.

\change{The emergent field gives rise to Landau levels that cause a topological Hall effect and also an orbital Hall effect, as shown previously for a skyrmion system~\cite{gobel2024topological} in analogy to a quantum Hall system~\cite{gobel2024OHE}.}
We have analyzed the Hall conductivities of a skyrmion texture in Ref.~\cite{gobel2024topological} but in a two-dimensional system. The results presented in Fig.~\ref{fig:skyrmion} differ from these because here we consider skyrmion tubes that extend infinitely along the $z$ direction. Landau levels that are rather flat in the $k_xk_y$ plane have dispersion along the $k_z$ direction because of the hopping of electrons along the skyrmion tube direction. This leads, for example, to the different band width and the missing quantization in the topological Hall conductivity in Fig.~\ref{fig:skyrmion}(d) in comparison to Ref.~\cite{gobel2024topological}. 

In summary, a skyrmion tube exhibits a similar orbital Hall conductivity $\sigma_{xy}^{L_z}$ as a hopfion but a hopfion exhibits an additional $\sigma_{yz}^{L_x}$ and $\sigma_{zx}^{L_y}$ orbital Hall conductivity. A skyrmion on the other side exhibits a sizable topological Hall response in the $\sigma_{xy}$ tensor element which is negligible for hopfions. Therefore, skyrmions and hopfions can be nicely distinguished in transport experiments.

\subsubsection{Bimeron tube}

The cross section of the hopfion torus consists of two in-plane skyrmions, also called bimerons. In fact, a hopfion is just a bimeron tube that has been twisted and closed to form a torus. Furthermore, a bimeron is topologically equivalent to a skyrmion, so it is natural to analyze the Hall transport of bimerons tubes next. The tube is chosen to extend along $x$ to resemble one of the torus plane directions.

\begin{figure}[t!]
    \centering
    \includegraphics[width=\columnwidth]{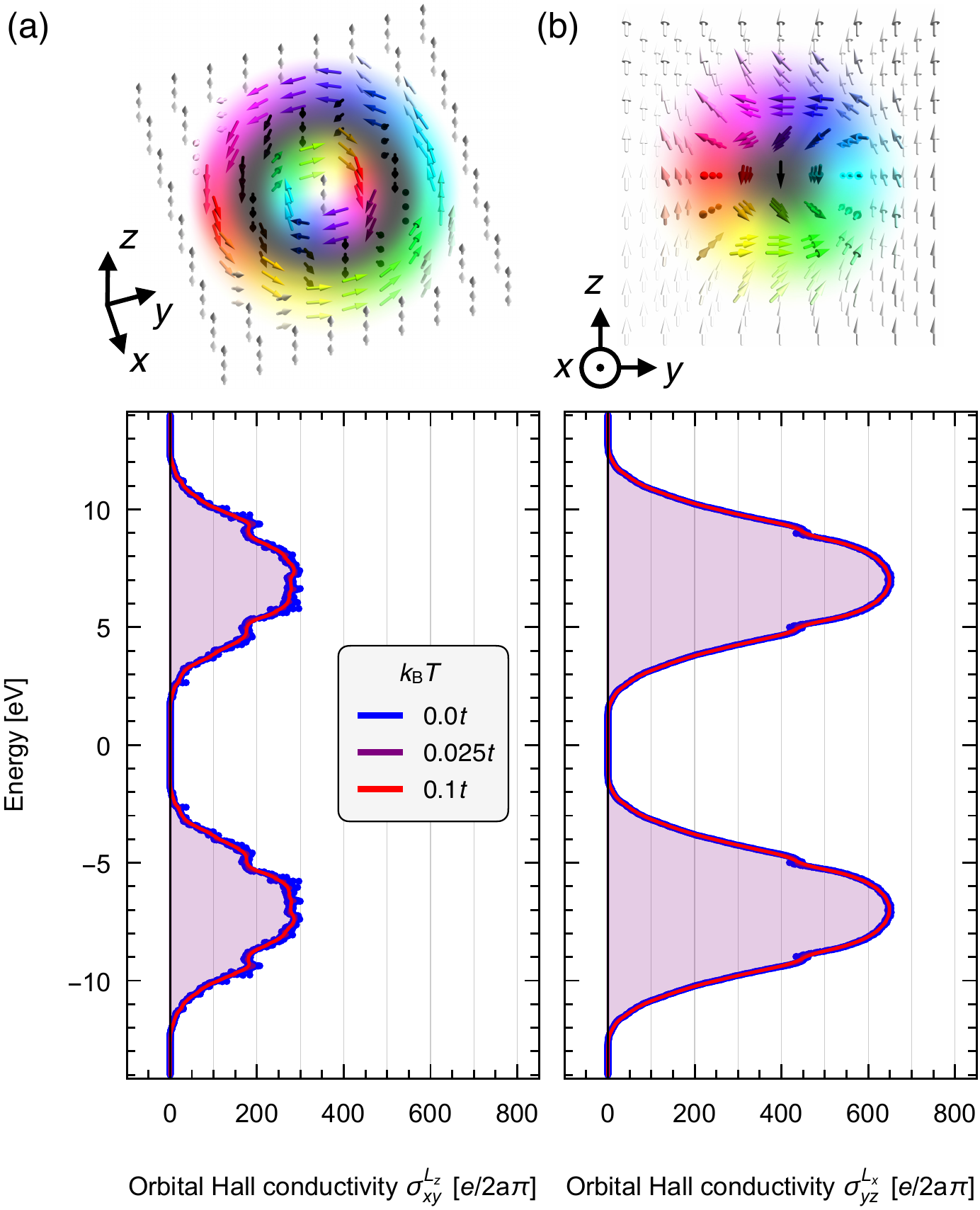}
    \caption{Orbital Hall effect caused by two-dimensional textures related to a hopfion.  (a) Top: Magnetic texture of a magnetic skyrmionium tube. The color indicates the orientation of the magnetic moments as in Fig.~\ref{fig:overview}. Bottom: Orbital Hall conductivity $\sigma_{xy}^{L_z}$ as a function of energy. The color of the curve indicates the temperature; see legend. (b) Calculations for a magnetic bimeron tube. Here, the tensor element $\sigma_{yz}^{L_x}$ is non-zero. The Hund's coupling is characterized by $m=7t$.}
    \label{fig:comparison}
\end{figure}

The orbital Hall conductivity $\sigma_{yz}^{L_x}$ is shown in Fig.~\ref{fig:comparison}(b) as a function of energy. It is exactly the same as the orbital Hall conductivity $\sigma_{xy}^{L_z}$ of a skyrmion tube that is shown in Fig.~\ref{fig:skyrmion}(c) and was discussed before. The reason is that we have constructed the bimeron such that it has the same profile as the skyrmion. The two textures can be transformed from one to another by spin rotations of $90^\circ$. Since the emergent field is invariant under collective spin rotations, bimerons and skyrmions exhibit the same emergent field. The only difference is that we have constructed the tubes along different directions ($z$ for the skyrmion and $x$ for the bimeron), which is why the two textures exhibit different orbital conductivity tensor elements.

Any cross section of a hopfion that contains the $z$ axis consists of two bimerons with opposite emergent fields. As we have shown now, such bimerons give rise to an orbital Hall conductivity along the bimeron plane. The response is the same for both bimerons, as the orbital Hall effect is insensitive to the sign of the emergent field. Each bimeron also exhibits a topological Hall conductivity $\sigma_{yz}$. However, it is exactly opposite for the two bimerons and cancels in total due to the opposite emergent fields.

In summary, a pair of oppositely magnetized bimeron tubes exhibits only an orbital Hall effect in the plane perpendicular to the tube direction. This explains why a hopfion exhibits orbital conductivity tensor elements $\sigma_{yz}^{L_x}$, $\sigma_{zy}^{L_x}$, $\sigma_{zx}^{L_y}$ and $\sigma_{xz}^{L_y}$ and why it does not exhibit a sizable topological Hall effect along these planes. Even the energy-dependencies of the orbital Hall conductivities exhibit a similar shape [cf. Fig.~\ref{fig:hopfion7}(b) and Fig.~\ref{fig:comparison}(b)] but with a different maximum due to the different average emergent field $\braket{B_{\mathrm{em},x}^2}$ and $\braket{B_{\mathrm{em},y}^2}$ of the two textures.

\subsubsection{Skyrmionium tube}

If a hopfion is cut horizontally, the resulting texture is a skyrmionium. This two-dimensional texture consists of a skyrmion that is positioned at the center of another skyrmion with opposite topological charge. The total topological charge is compensated and therefore the average emergent field is zero. However, since the two subskyrmions exhibit out-of-plane emergent fields, a skyrmionium exhibits an average $\braket{B_{\mathrm{em},z}^2}$. Here we consider a skyrmionium tube where the two-dimensional skyrmionium texture has been extended trivially along the $z$ direction.

The skyrmionium tube exhibits an orbital Hall conductivity $\sigma_{xy}^{L_z}$ that is shown in Fig.~\ref{fig:comparison}(a). Due to the small size of the skyrmionium and the inhomogeneity of the emergent field, the curve is more noisy but in general has a similar energy dependence as all orbital Hall conductivities discussed before. 
This result for the skyrmionium allows us to understand why the hopfion exhibits an orbital Hall conductivity tensor element $\sigma_{xy}^{L_z}$; cf. Fig.~\ref{fig:hopfion7}(a). Both, the skyrmionium and the hopfion, do not exhibit a sizable topological Hall conductivity $\sigma_{xy}$ due to the compensated emergent field, on average.

\subsubsection{Summary}

In summary, the three-dimensional orbital Hall response of a hopfion can be understood by analyzing the two-dimensional textures that appear when the hopfion is cut along a two-dimensional plane: A cut along the $xy$ plane reveals a skyrmionium that exhibits only an orbital Hall conductivity $\sigma_{xy}^{L_z}$ and no corresponding topological Hall conductivity. A cut along the $yz$ plane reveals a pair of bimerons that only give rise to an orbital Hall conductivity $\sigma_{yz}^{L_x}$ and not to a corresponding topological Hall conductivity. An equivalent result for $\sigma_{zx}^{L_y}$ can be found for the pair of bimerons that appear when cutting the hopfion along the $zx$ plane.


\subsection{Approximating the orbital Hall conductivities} \label{sec:results_approximation}

As discussed in the previous sections, all non-zero orbital Hall conductivities of the various textures exhibit a similar energy dependence only with different magnitudes. The magnitude of $\sigma_{\beta\gamma}^{L_\alpha}$ is determined by the average emergent field component squared $\braket{B_{\mathrm{em},\alpha}^2}$. This is because the emergent field causes the transverse transport of electronic states and also causes their orbital polarization. In a two-current model, oppositely orbital polarized states move along opposite directions and give rise to a net orbital current that is independent of the sign of the emergent field. 

The recurring energy dependence is determined by the band structure of the underlying cubic lattice. We have already briefly identified the role of this band structure for the distribution of the electronic states in energy by discussing the density of states in Sec.~\ref{sec:results_transport} and we will do so in more detail in the following.

As we have established in a previous work on magnetic skyrmions~\cite{gobel2024topological}, the energy dependencies of the charge, spin and orbital Hall conductivities can be approximated by the density of states $n'_{d\mathrm{D}}(E)=\frac{1}{(2\pi)^d}\int \delta(E-E_{d\mathrm{D}}(\vec{k}))\,\mathrm{d}^dk$ and the carrier density $n_{d\mathrm{D}}(E)=\kappa(E)\int^E n'_{d\mathrm{D}}(E')\,\mathrm{d}E'$ calculated from the band structure of the underlying lattice $E_{d\mathrm{D}}(\vec{k})$. Here, $\kappa(E)=\pm 1$ is determined by the electronic character of the Fermi line: electron-like ($+1$) vs hole-like ($-1$). In that reference, a two-dimensional system, $d=2$, in the $xy$ plane with an emergent field along $z$ had been considered, and we found that the orbital Hall conductivity can be approximated as
\begin{align}
     \sigma_{xy,\mathrm{approx,2D}}^{L_z}(E)\propto \frac{[n_\mathrm{2D}(E)]^2}{n'_\mathrm{2D}(E)}\frac{1}{\braket{B_{\mathrm{em},z}^2}}.
\end{align}
Here, we generalize these findings of the two-dimensional system to explain the results of three-dimensional systems, $d=3$, that were shown in Sec.~\ref{sec:results_transport}. As an example, we discuss $\sigma_{xy}^{L_z}$ that is determined by $\braket{B_{\mathrm{em},z}^2}$. The other tensor elements can be calculated analogously by rotating the coordinate system.  

\begin{figure}[t!]
    \centering
    \includegraphics[width=\columnwidth]{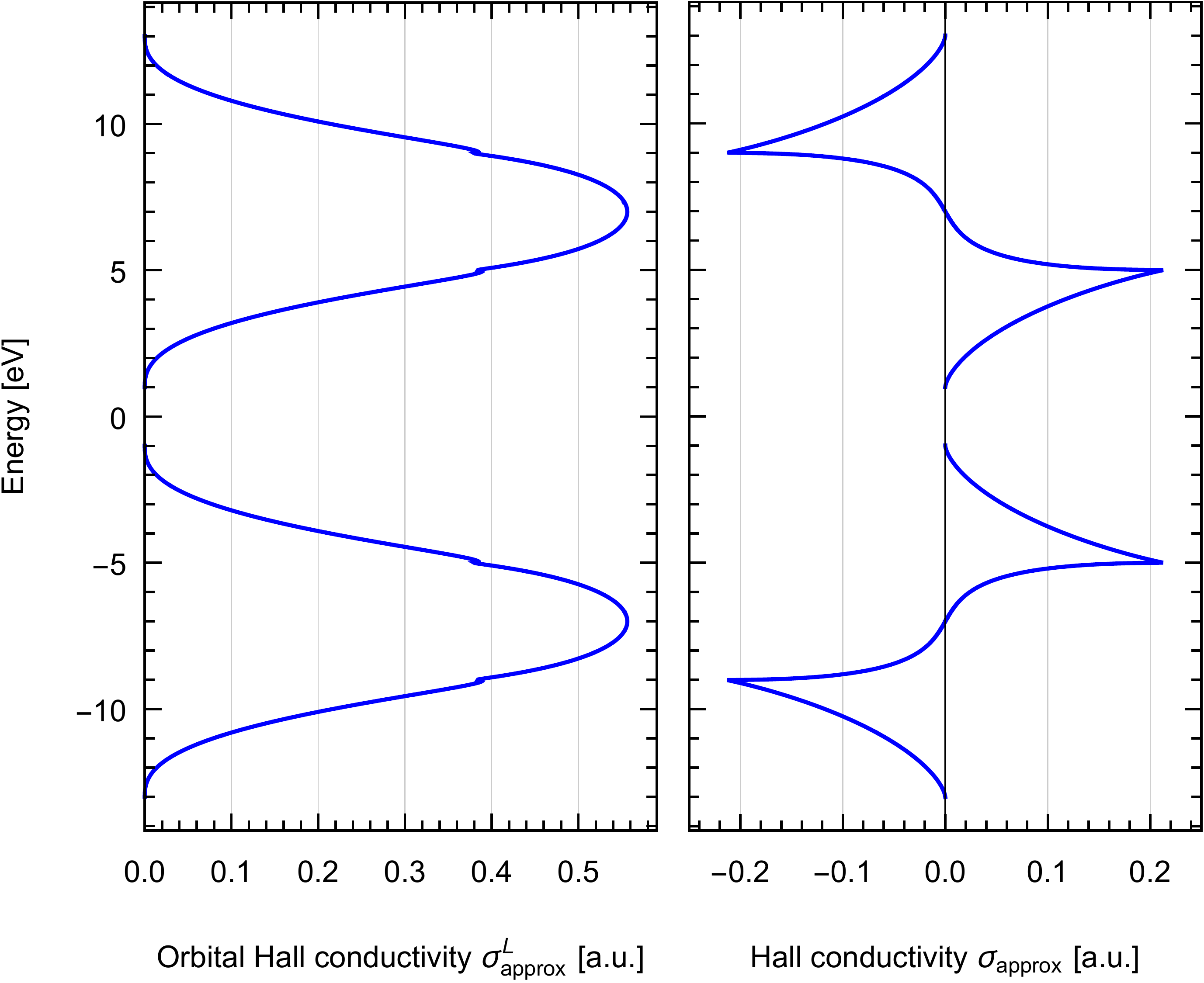}
    \caption{Approximation of energy-dependent Hall conductivities for textures with a finite emergent field in the strong-coupling limit. (a) Orbital Hall conductivity $\sigma_\mathrm{approx}^L$ with the orbital angular momentum $L$ along the emergent field and the electric field and the transverse orbital current in the plane perpendicular to it. (b) Topological Hall conductivity $\sigma_\mathrm{approx}$ in the plane perpendicular to the emergent field. The Hund's coupling is $m=7t$ so that the conduction electrons' spins can be assumed to align with the magnetic texture.}
    \label{fig:approximation}
\end{figure}

When transitioning from the two-dimensional system to the three-dimensional system, the band structure of the underlying lattice changes from $E_\mathrm{2D}(k_x,k_y)=-2t\left[\cos(k_xa)+\cos(k_ya)\right]$ to $E_\mathrm{3D}(k_x,k_y,k_z)=-2t\left[\cos(k_xa)+\cos(k_ya)+\cos(k_za)\right]$, so in this particular case $E_\mathrm{3D}(k_x,k_y,k_z)=E_\mathrm{2D}(k_x,k_y) -2t\cos(k_za)$. However, because the transverse transport occurs in the plane perpendicular to $\vec{B}_{\mathrm{em}}$ (here the $xy$ plane), we cannot simply replace the density of states and carrier density of the two-dimensional system with those of the three-dimensional system. Instead, we consider a modified density of states, the $k_z$-resolved density of states
\begin{align}
    n'(E,k_z)=\frac{1}{(2\pi)^3}\int \delta(E-E_{3\mathrm{D}}(k_x,k_y,k_z))\,\mathrm{d}k_x\mathrm{d}k_y
\end{align}
and carrier density
\begin{align}
    n(E,k_z)=\kappa(E,k_z)\int^E n'(E',k_z)\,\mathrm{d}E'
\end{align}
and integrate over the $\vec{k}$ direction along $\vec{B}_{\mathrm{em}}$ (here along $k_z$). The resulting orbital Hall conductivity
\begin{align}
     \sigma_{xy,\mathrm{approx,3D}}^{L_z}(E)\propto \frac{1}{\braket{B_{\mathrm{em},z}^2}}\int \frac{[n(E,k_z)]^2}{n'(E,k_z)}\,\mathrm{d}k_z.
\end{align}
is shown in Fig.~\ref{fig:approximation}(a) and resembles the previously calculated orbital Hall conductivities of the various non-collinear magnetic textures well; cf. Figs.~\ref{fig:hopfion7}(a,b), \ref{fig:skyrmion}(c), \ref{fig:comparison}(a,b). 

The results of the full quantum-mechanical calculations are especially well reproduced for the skyrmion tube and the bimeron tube; cf. Figs.~\ref{fig:skyrmion}(c), \ref{fig:comparison}(b). For these textures, the emergent field is rather smooth and does not change along the tube direction, so that the assumptions of the approximation are well fulfilled: The electronic states are distributed according to the density of states of the underlying lattice and the orbital Berry curvature does not change much along the $\vec{k}$ direction that is along the direction perpendicular to the Hall transport. As another example, the hopfion's $\sigma_{xy}^{L_z}$ component [Fig.~\ref{fig:hopfion7}(a)] is also nicely approximated but there are small deviations. These occur because the emergent field is less smooth [Figs.~\ref{fig:overview}(b)] and the orbital Berry curvature $\Omega_{xy}^{L_z}$ varies slightly along $k_z$ [Figs.~\ref{fig:bandstructure}(b)].

Overall, we have achieved a good semi-classical understanding of the orbital Hall effect in these three-dimensional systems. The magnitude of $\sigma_{\beta\gamma}^{L_\alpha}$ is determined by the average emergent field $\braket{B_{\mathrm{em},\alpha}^2}$ and the energy dependence can be reconstructed based on the zero-field band structure of the underlying lattice $E_{\mathrm{3D}}(\vec{k})$. For the textures with a finite topological charge $N_\mathrm{Sk}$ (in this paper the skyrmion tube and the bimeron tube), we can also approximate the topolgical Hall conductivity and the spin Hall conductivity as
\begin{align}
    \sigma_{xy,\mathrm{approx,3D}}(E)&\propto \pm \frac{1}{\braket{B_{\mathrm{em},z}}} \int n(E,k_z)\,\mathrm{d}k_z,\\
     \sigma_{xy,\mathrm{approx,3D}}^{S_z}(E)&\propto \,\,\,\,\frac{1}{\braket{B_{\mathrm{em},z}}}\int n(E,k_z)\,\mathrm{d}k_z,
\end{align}
by generalizing the two-dimensional formulas from Ref.~\cite{gobel2024topological} in the same way as for the orbital Hall conductivity before.
The result for the topological Hall conductivity is shown in Fig.~\ref{fig:approximation}(b) and reproduces the results of the full calculation very well; cf. Fig.~\ref{fig:skyrmion}(d).

\change{Our analysis allows the understanding of the functional dependencies of the Hall conductivities presented throughout this paper. For the following discussion we consider the states with negative energy in Fig.~\ref{fig:approximation}. The topological Hall conductivity, shown in panel (b), characterizes charge transport. The energetically lower states are electron-like and the higher states of each block are hole-like; the effective mass changes sign and the states are affected oppositely by the (emergent) magnetic field. Consequently, $\sigma_\mathrm{approx}(E)$ exhibits a sign change and has two extrema. However, the orbital Hall conductivity, shown in panel (a), characterizes transport of orbital angular momentum. The orbital angular momentum of a Bloch state is also dependent on the effective mass. Therefore, electron-like and hole-like states behave equally in terms of the orbital Hall transport. Consequently, $\sigma_\mathrm{approx}^L(E)$ does not change sign and exhibits only one extremum. By the same arguments, we can understand that the orbital Hall conductivity is an even function of the (emergent) magnetic field, while the topological Hall effect is odd~\cite{gobel2024OHE,gobel2024topological}.}



\section{Conclusion} \label{sec:conclusion}

In summary, we have shown that a hopfion exhibits a three-dimensional topological orbital Hall effect. The emergent field itself is large within the hopfion but is compensated on a global level which is why the charge and spin Hall conductivities are almost compensated. The square of the emergent field, however, is finite on average and causes the orbital Hall effect, which can be seen as a unique signature of the hopfion phase: Since the hopfion is an innately three-dimensional texture, it exhibits multiple tensor elements $\sigma_{xy}^{L_z}$, $\sigma_{yz}^{L_x}$ and $\sigma_{zx}^{L_y}$ [cf. Fig.~\ref{fig:hopfion7}], whereas a skyrmion tube, for example, is continued trivially along the tube direction and exhibits only an orbital Hall transport within the skyrmion plane. 

Two-dimensional textures, such as skyrmions, are characterized by the topological charge $N_{\mathrm{Sk}}$ that determines their topological Hall transport. The hopfion is a three-dimensional magnetic texture that is topologically characterized by the Hopf index. However, so far, the significance of this topological invariant for a hopfion's emergent electrodynamics has not been known. Here we have shown that the three-dimensional orbital Hall response of a hopfion can be seen as a signature of the Hopf index. All results presented in this paper can be interpreted based on the emergent field of the texture which, in turn, determines its topological invariant.

Our findings present a reliable hallmark for identifying hopfions in experiments. This was difficult so far, because the commonly used real-space techniques like LTEM penetrate the sample and result in a projected image of the texture that can hardly be distinguished from a skyrmionium. Measuring the orbital response of hopfions presents an alternative to difficult and expensive holography and tomography measurements~\cite{wolf2019holographic,midgley2009electron,hierro2020revealing,wolf2022unveiling,seki2022direct,yasin2024bloch,winterott2025unlocking,gubbiotti20242025}. Furthermore, hopfions can serve as generators of large orbital currents that can give rise to large torques which are needed for spin-orbitronic devices.


\paragraph*{Acknowledgements ---}
This work was supported by the EIC Pathfinder OPEN grant 101129641 ``Orbital Engineering for Innovative Electronics''. 

\paragraph*{Data Availability ---}
The data that support the findings of this article are openly available~\cite{data}.

\bibliography{short,MyLibrary}

\end{document}